# Cepheid Variables in the Flared Outer Disk of our Galaxy

Michael W. Feast, John W. Menzies, Noriyuki Matsunaga & Patricia A. Whitelock

**Flaring and warping of the disk of the Milky Way have been inferred from observations of atomic hydrogen[1,2] but stars associated with flaring have not hitherto been reported. In the area beyond the Galactic centre the stars are largely hidden from view by dust, and the kinematic distances of the gas cannot be estimated. Thirty-two possible Cepheid stars (young pulsating variable stars) in the direction of the Galactic bulge were recently identified[3]. With their well-calibrated period–luminosity relationships, Cepheid stars are useful distance indicators[4]. When observations of these stars are made in two colours, so that their distance and reddening can be determined simultaneously, the problems of dust obscuration are minimized. Here we report that five of the candidates are classical Cepheid stars. These five stars are distributed from approximately one to two kiloparsecs above and below the plane of the Galaxy, at radial distances of about 13 to 22 kiloparsecs from the centre. The presence of these relatively young (less than 130 million years old) stars so far from the Galactic plane is puzzling, unless they are in the flared outer disk. If so, they may be associated with the outer molecular arm[5].**

We derived the distances for the five Cepheids from near-infrared photometry obtained with the Infrared Survey Facility (IRSF) and we used radial velocities from the Southern African Large Telescope (SALT) to determine the kinematics (see Methods)—both telescopes are at the South African Astronomical Observatory (SAAO), Sutherland, in South Africa. From these data we were able to ascertain the population to which the Cepheids belong. The other 27 Cepheid candidates are either better assigned to a different class (such as anomalous Cepheids) or else their classification as classical Cepheids is uncertain.

Table 1 lists the derived distances and various other parameters for the Cepheids. They are at about the distance and position at which a stream associated with the Sagittarius (Sgr) dwarf galaxy crosses the plane[6], but the low radial velocity (mean heliocentric radial velocity after correction for the effects of stellar pulsation of $V_R = 4 \pm 8$ km s$^{-1}$, see Table 1) is completely different from that expected for members of the Sgr dwarf stream (about 150 km s$^{-1}$)[6,7] and the Cepheids are clearly Galactic. They cannot be in the Galactic bulge because their distances from the centre put them far beyond the bulge and the velocity dispersion of the five stars, $16 \pm 5$ km s$^{-1}$ (much of which is observational), is much smaller than expected for bulge objects (>60 km s$^{-1}$)[8]. Furthermore, these short-period Cepheids will be relatively young (about 100 million years (Myr) old), and, although there is a young component, including Cepheids[9], in the innermost regions of the bulge, the bulk of the population is old (about 10 billion years (Gyr) old)[8]. Figure 1 shows the positions of the five stars in comparison to catalogued Cepheids. The various sources of uncertainty for the distances of the Cepheids are discussed in the Methods, but the reddening law and reddening corrections presented the biggest challenge and are the primary contributors to the error bars shown in the figure.

**Table 1: Data for individual Cepheids**

| OGLE # | $l$ (deg) | $b$ (deg) | $D$ (kpc) | $z$ (kpc) | $R$ (kpc) | $V_R$ (km s$^{-1}$) | $\rho$ (km s$^{-1}$) | $P$ (day) |
|---|---|---|---|---|---|---|---|---|
| 01 | -0.03 | 2.94 | 28.4 | 1.5 | 19.9 | -12 | -3 | 2.598 |
| 02 | 4.57 | 4.85 | 25.2 | 2.1 | 16.6 | 27 | 50 | 2.026 |
| 03 | 4.35 | 2.89 | 25.4 | 1.3 | 16.9 | 19 | 24 | 1.236 |
| 05 | 5.38 | 2.34 | 25.2 | 1.0 | 16.8 | 23 | 28 | 3.796 |
| 32 | 6.89 | -3.89 | 29.5 | -2.0 | 21.4 | 10 | 15 | 3.736 |

OGLE names are OGLE-BLG-CEP-#, $l$ and $b$ are the galactic coordinates; $D$, $z$ and $R$ are the distances from the sun, the Galactic plane and the perpendicular distance from the axis of Galactic rotation (assuming the Sun-Galactic centre distance is 8.5 kpc), respectively; $V_R$ and $\rho$ are the heliocentric radial velocity and the radial velocity after correction for solar motion and Galactic rotation, respectively; and $P$ the pulsation period.

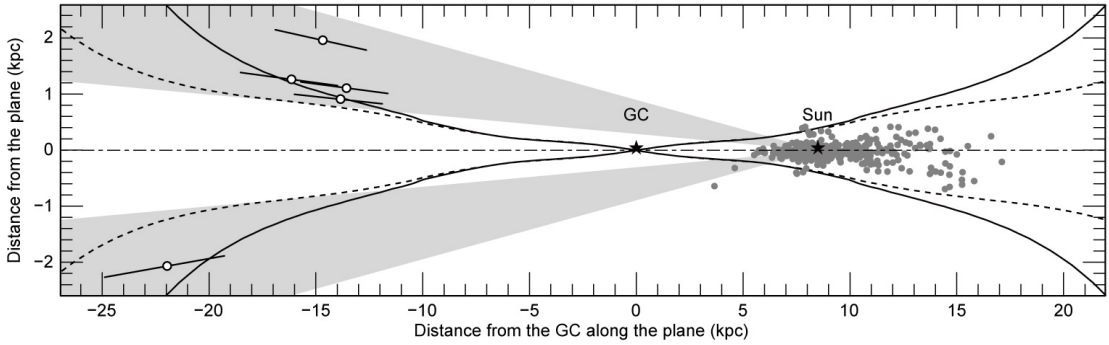

**Figure 1: Schematic of the Galaxy.**
The positions of the Cepheids (open circles with assumed maximum uncertainties of ±0.2 mag) are compared to the location of the H I gas. The solid and dashed curves are model fits, S and N1, respectively, from ref. 1 at three times the HWHM above and below the Galactic plane. We note that figures 1 and 2 of ref. 2 show the H I flare in the relevant region extending up to about 2 kpc. The dark grey points are previously known Galactic Cepheids[10] and the approximate regions surveyed by OGLE (2 < |b| < 6) are shown in light grey on either side of the plane. The positions of the Sun and Galactic centre are indicated by the star symbols.

There is almost no information on gas or stars in the Galactic disk immediately behind (Galactic longitude $l \pm 15°$) the Galactic centre. The atomic hydrogen observations[2] on either side of the centre, but away from the central region itself, suggest that the gaseous disk of the Milky Way at $l \approx 0$ is not warped but shows a marked flaring at Galactocentric radii ($R$, the distance from a star to the centre of the Galaxy) of 15 kiloparsecs (kpc) and more; we note that the details are model dependent. The thickness of the gaseous disk[1,2] increases from 60 parsecs (pc) half-width at half-maximum (HWHM) at $R = 4$ kpc to 2.7 kpc at $R = 30$ kpc and, especially at positive Galactic longitudes[1], there is a marked increase from about 0.4 kpc at $R = 15$ kpc to about 1.0 kpc at $R = 20$ kpc.

Therefore we found the Cepheids at exactly the distance predicted for this increase in disk thickness, as can be seen in Fig. 1. The absence of Cepheids nearer the Sun is consistent with the lower HWHM in these regions, whereas the absence of more distant Cepheids is partly due to the decreasing density at larger distances from the centre and partly the consequence of the Optical Gravitational Lensing Experiment (OGLE) observational cut-off. So the relatively narrow range of distances is consistent with our hypothesis that these stars are in the flared disk. In the Methods we also show that the numbers of Cepheids observed is consistent with expectations from a flared disk.

Cepheids are usually associated with spiral arms and the distances of these five are similar to that expected for the far outer molecular spiral arm of the Galaxy[5] where it passes behind the central region of the Galaxy; the HWHM of this arm may be only about 0.6 kpc, in which case the Cepheids would be on its periphery. However, we note that distances and thickness computed for this arm depend sensitively on the model adopted and are therefore uncertain[2,5].

It is instructive to examine why the outer regions of a galactic disk flare. In the inner parts of a galactic disk the gravitational force $k(z)$ at height $z$ perpendicular to the galactic plane is dominated by the strong concentration of stars there. As we move to greater galactocentric radii, however, the concentration of stars drops dramatically, $k(z)$ decreases and is increasingly dominated by the effects of dark matter. The flaring of the gas layer in the outer parts of our own and other galaxies has been attributed to this, and observations can in principle be used to study the distribution of dark matter in the halo of galaxies[11]. Studies of the flaring of HI gas in our Galaxy[1] suggest that in addition to an isothermal dark halo of $1.8 \times 10^{12}$ $M_\odot$ (where $M_\odot$ is the mass of the Sun), there is a self-gravitating exponential dark-matter disk ($1.8–2.4 \times 10^{11}$ $M_\odot$) as well as a dark-matter ring (13 kpc $< R <$ 18.5 kpc and $2.2–2.8 \times 10^{10}$ $M_\odot$), which may represent the remains of a cannibalized dwarf galaxy[1]. The most serious uncertainty in using gas as a tracer of the gravitational field arises from the need to adopt a model to derive the gas distribution. It is therefore highly desirable that the gravitational field in the outer Galaxy be investigated using young stars for which good distance estimates can be made. Classical Cepheid variables are by far the best stars for this purpose.

Studies of diffuse groups of B stars[12], which are even younger than Cepheids, are also consistent with a Galactic disk extending 15 kpc and 20 kpc from the centre, at Galactic latitude $b = -4°$ and $-7°$, respectively. These stars are in the third Galactic quadrant near the place where the warp forces the Galactic plane to its greatest negative displacement from $b = 0°$. So although these young stars are displaced from $b = 0°$, they are in the local Galactic plane, and therefore tell us nothing about a flare.

The collection of stars now known as the 'Monoceros ring' has been interpreted as evidence for a warped disk[13], or alternatively as the remnant of a dwarf galaxy cannibalized by the Milky Way[14]. It is perhaps curious that the Cepheids discussed above are at the distance from the Galactic centre that one would expect the Monoceros ring to be, if indeed it were a complete circular ring around the Galaxy. The stellar population that makes up this so-called ring is generally considered to be old (>1 Gyr) and therefore different from the Cepheids (although there have been suggestions of an association with spiral arms[11]). Models[15] indicate that the ages of the youngest Cepheids discussed here are less than 130 Myr. The disputed origin of the Monoceros ring[16] is beyond the scope of this Letter. Nevertheless, we note that simulations that suggest that the ring is a consequence of the interaction of the Sgr dwarf galaxy with the Milky Way[17] do not predict any significant density of stars in the ring at the distance of the Cepheids under discussion.

Clearly, these Cepheids are just the tip of the iceberg. Further work on these stars and other 'standard candles' in the outer Galaxy will present new opportunities to probe the gravitational field and therefore the distribution of dark matter in the outer parts of our Galaxy.

**Methods**
In the following, we describe how the stars were identified as classical Cepheids by comparison with similar stars in the LMC[18]. We then go on to derive distances, taking into account the well known abnormal reddening law towards the Galactic centre[19].

**Identifying classical Cepheids**
A problem when studying Cepheids is that it is not always easy from available photometry to distinguish classical Cepheids (type I) from other objects, for example, anomalous or type II Cepheids (BL Her stars, W Vir stars). This issue has been discussed in the context of distant Cepheids towards the anti-centre[20] (that is, the direction in the Galaxy that is opposite from the centre, viewed from our perspective). In the interior region of the Galaxy, and particularly in the direction of the bulge, this is likely to be a significant problem. Fortunately it is possible to distinguish between some classes of stars using the Fourier coefficients of their light curves, and these are listed for the OGLE[3] Cepheids towards the bulge.

The main Fourier parameters for the *I*-band light curves of the five stars discussed here (Extended Data Table 1) can be compared with plots of the Fourier coefficient ratios $R_{21}$, $R_{31}$ and phase differences $\phi_{21}$, $\phi_{31}$ (where the subscripts denote the order of the cosine curve fit) against period for various classes of variable star in the LMC[18] and this enables us to classify these five securely as classical Cepheids. Other possible Cepheids in the OGLE bulge catalogue have characteristics that suggest they belong to the anomalous Cepheid class, are possible type II Cepheids or else their classification is doubtful.

**Extended Data Table 1: Fourier coefficients for the *I*-band light curves of the Cepheids**

| OGLE # | Log P | $R_{21}$ | $\varphi_{21}$ | $R_{31}$ | $\varphi_{31}$ |
|---|---|---|---|---|---|
| 01 | 0.414 | 0.488 | 4.458 | 0.302 | 2.807 |
| 02 | 0.307 | 0.522 | 4.263 | 0.354 | 2.444 |
| 03 | 0.092 | 0.099 | 3.603 | 0.157 | 1.706 |
| 05 | 0.579 | 0.443 | 4.753 | 0.212 | 3.242 |
| 32 | 0.572 | 0.436 | 4.556 | 0.232 | 2.791 |

The amplitude (*A*) ratios $R_{n1} = A_n/A_1$ and phase differences $\phi_{n1} = \phi_n - n\phi_1$ are listed, where $A_n$ and $\phi_n$ are parameters of the truncated Fourier series fitted to the photometric data[3,17]. The subscripts refer to the order of the fit, so that $n = 1$ is the fundamental, $n = 2$ is the first harmonic and so on. *P* is the pulsation period in days.

**Photometry**

The infrared photometry (Extended Data Table 2) was carried out using the 1.4-m IRSF and the SIRIUS camera at Sutherland[21]. Each of the targets was observed once on 2012 May 6 (Universal time, UT) with an exposure time of 25s (5s times five exposures at dithered positions). The photometry was extracted using the Image Reduction and Analysis Facility (IRAF) package DAOPHOT (http://iraf.noao.edu) and standardized by comparison with nearby stars from the 2MASS point source catalogue[22]. The uncertainties for the brightest and faintest of the Cepheids range from 0.02–0.07 mag at $J$, 0.02–0.03 mag at $H$ and 0.02–0.04 mag at $K_S$, respectively. These are significantly less than the uncertainties on the 2MASS measures, where they exist, for the same sources. We use these single-epoch $J$, $H$ and $K_S$ measurements to estimate the distance, noting that the near-infrared amplitudes of these short-period stars will be small (<0.1 mag; ref. 23).

**Extended Data Table 2: Photometry of the Cepheids**

| OGLE # | 01 | 02 | 03 | 05 | 32 |
|---|---|---|---|---|---|
| $V$ | 20.800 | 17.482 | 18.207 | 17.675 | 16.731 |
| $I$ | 17.382 | 15.682 | 16.390 | 15.374 | 15.047 |
| $J$ | 15.28 | 14.34 | 15.25 | 13.67 | 14.04 |
| $H$ | 14.03 | 13.79 | 14.61 | 13.03 | 13.42 |
| $K_S$ | 13.79 | 13.63 | 14.34 | 12.74 | 13.33 |
| JD-2456053 | 0.61250 | 0.61389 | 0.61458 | 0.61667 | 0.64514 |
| $(\mu_0)_{VI}$ | 17.24 | 17.01 | 17.02 | 17.01 | 17.35 |
| $(\mu_0)_{JK}$ | 16.96 | 16.83 | 16.72 | 16.74 | 17.42 |
| $A_I$ | 3.12 | 1.33 | 1.39 | 1.83 | 1.14 |
| $A_K$ | 0.57 | 0.19 | 0.31 | 0.28 | 0.17 |

For each star the OGLE mean $V$ and $I$ magnitudes and the single epoch IRSF $J$, $H$ and $K_S$ magnitudes for observations made on the given Julian date (JD) are listed. $(\mu_0)_{VI}$ is the reddening-corrected distance modulus calculated from equations (1) and (2) and $(\mu_0)_{JK}$ is the reddening-corrected distance modulus calculated from equations (3) and (4). $A_I$ and $A_K$ are the interstellar extinction values at $I$ and $K_S$ respectively, calculated simultaneously with the distance moduli.

**Distances and interstellar absorptions**

In general there are severe problems in dealing with observations of distant stars in the Galactic plane close to or beyond the centre because of the large and uncertain amounts of interstellar extinction in these directions. Cepheids offer an important advantage in this regard in that distances can be derived from relations that allow the reddening and the distance to be determined together and unambiguously when observations in two colours are available—for example, $V$ and $I$ or $J$ and $K_S$—provided the reddening law is known. Recent work[19] has indicated that the law of reddening is different towards the Galactic bulge from that adopted elsewhere[24] and here we use:

$A_K = (0.494 \pm 0.006) E_{(J-K)}$

from ref. 19 and

$A_I = (1.125 \pm 0.09) E_{(V-I)}$,

which was found by the same method[25]. It should be noted that the relation in $V$ and $I$ may be somewhat more complex than the one given[25].

Adopting period–luminosity relations in $V$ and $I$ (as derived[26] by the OGLE group) and $J$ and $K_S$ (derived[27] for Cepheids with 0.4<log$P$<1.0) from the LMC together with an LMC distance modulus of 18.5 mag and interstellar extinction values[28] of $A_V$ = 0.22 mag, $A_I$ = 0.13 mag, $A_J$ = 0.06 mag and $A_K$ = 0.02 mag for the LMC direction, we then have the distance modulus $\mu_0$ for a Cepheid with a pulsation period $P$:

$$\mu_0 + A_V = V + 2.762 \log P + 1.190, \qquad (1)$$
$$\mu_0 + A_I = I + 2.959 \log P + 1.751 \qquad (2)$$

and

$$\mu_0 + A_J = J + 3.138 \log P + 2.109, \qquad (3)$$
$$\mu_0 + A_K = K_S + 3.284 \log P + 2.383. \qquad (4)$$

Combining these pairs of equations with the reddening law given above leads to the two estimates of the distance modulus, $(\mu_0)_{VI}$ and $(\mu_0)_{JK}$—derived, respectively, from equations (1) and (2) and equations (3) and (4)—and the interstellar extinction corrections, $A_I$ and $A_K$ (Extended Data Table 2).

**Uncertainties in the distances**
The LMC period–luminosity relations that we used are well defined[4]. Their absolute calibration is based on the LMC distance modulus, which has been determined in a number of ways. The uncertainty in the adopted value (about 0.04 mag or 2%) is negligible for our discussion. The mean OGLE $VI$ magnitudes derived from their extensive observations have negligible error. Because of the small pulsation amplitudes of the Cepheids in the infrared region of the spectrum, the error on our $J$ and $K_S$ magnitudes is ≤0.05 mag.

Possible metallicity effects on the period–luminosity relations have been much discussed[4]. Nothing is known about the metallicities of Cepheids behind the Galactic centre, but those of Cepheids in the outer disk of the Galaxy in the general direction of the anti-centre[29], and at comparable distances from the centre to those discussed here, have a mean logarithmic iron-to-hydrogen ratio [Fe/H] = −0.60 ± 0.12, which is intermediate between those of the LMC and the Small Magellanic Cloud[30]. The difference in distance between the Small Magellanic Cloud and the LMC derived from $J$ and $K_S$ observations of Cepheids[31] agrees with values measured in other ways, without the application of metallicity corrections. Furthermore, Hubble Space Telescope parallaxes of Galactic Cepheids[32] (with [Fe/H] ≈ 0) agree with the LMC modulus adopted without the application of any metallicity corrections (they give 18.52 ± 0.03 mag from $V$ and $I$ and 18.47 ± 0.03 mag from $K_S$). The various factors indicate that any residual metallicity effects on the distances derived for these Cepheids will be very small.

A potential source of uncertainty is in the width of the period–luminosity relations. This width is due to the fact that, at a given period, a Cepheid brighter than the average is also bluer. This leads to the smaller-than-average $V$ being compensated by a lower-than-average derived apparent absorption. It is clear that the uncertainty in the modulus due to the spread in colour at a given period is[4]:
$$\sigma(\mu_0) = (\beta_1 - \beta_2)\sigma(V-I)_0, \qquad (5)$$
where $\beta_1$ is the colour coefficient of a (nearly dispersionless) period–luminosity–colour relation in $(V, I)$ and $\beta_2$ is the ratio of total to selective absorption. For the Cardelli[24] law of reddening, which is often used, $\beta_1 \approx \beta_2$. Thus, any uncertainty due to the width of the period–luminosity relation in our case comes from the change in $\beta_2$ for the bulge, which is 0.33. The scatter in $V - I$ at a given period[33] is 0.08, which would result in $\sigma(\mu_0) = 0.03$ in equation (5). In the infrared, the widths of the period–luminosity relations are lower and will not introduce significant uncertainty.

Interstellar reddening is a source of error and, as pointed out above, the evidence points to an abnormal reddening law in the direction of these stars. The uncertainty in this reddening law in $JK_S$ is small; this, together with the low extinction in the infrared, leads to only a small uncertainty in the distance modulus (0.003 mag for the most heavily reddened star). Even if, contrary to the evidence, we used the Cardelli[24] law of reddening, the change in distance moduli would not affect our conclusions. In that case, the modulus of our most reddened star would decrease by 0.28 mag (a change of distance from 24.4 kpc to 21.4 kpc) and the moduli of the other stars would decrease by an average of 0.12 mag (1.25 kpc). Owing to the greater absorption in $V$ and $I$ and the greater uncertainty in the reddening law, the uncertainties in the derived distances are greater. The uncertainty in the reddening coefficient leads to an uncertainty of 0.25 mag in the modulus of the most heavily reddened star and a mean of 0.12 mag in the other cases. If, contrary to expectations, a Cardelli reddening law had been adopted, the modulus of the most heavily absorbed star would have decreased by 0.90 mag and those of the others by a mean of 0.42 mag. Clearly, reddening uncertainties in $VI$ are much more important than in $JK$.

**Summary of adopted distances and their uncertainties**
In the main paper we adopt the distances derived from the $J$ and $K_S$ magnitudes (see Table 1), because they are the more accurate values. The above discussion indicates that the errors in those distance moduli are: 0.04 mag from the absolute calibration; ≤0.05 mag due to the pulsation amplitude; and negligible amounts from the period–luminosity relation width, metallicity effects and uncertainties in the Nishiyama reddening law. If the Cardelli reddening law were applied to these stars their moduli would be reduced by a mean of 0.15 mag, but a change this large seems to be ruled out by observations. The systematic uncertainties overwhelm the rather small statistical errors, so we do not attempt to assign individual errors to distances. We consider 0.2 mag to be a very conservative estimate of the total error of an individual modulus (random plus systematic) and this is what is illustrated in Fig. 1, but we fully expect the errors to be less than this. In the case of the moduli from $V$ and $I$, the main uncertainty is from the coefficient of the reddening law and complications in deriving this have been noted[25]. We simply mention here that with the adopted law the $VI$ moduli are 0.20 mag larger than the $JK_S$ values adopted, whereas with a Cardelli law they are 0.31 mag smaller, suggesting that a less extreme variation from the Cardelli law applies to these stars.

**Radial velocities**
Our spectra (Extended Data Table 3) were obtained with the Robert Stobie spectrograph on the SALT. A volume phase holographic grating of 1,300 lines mm$^{-1}$ was used to cover the wavelength range 7,800–9,600 Å, putting the Ca II triplet on the middle charge-coupled device of the detector. The resolution is 3.4 Å with a projected slit width of 1.5 arcsec.

**Extended Data Table 3: Journal of spectroscopic observations**

| Object | HJD − 2450000.0 | Phase/($V_R$) |
|---|---|---|
| OGLE-BLG-CEP-01 | 6433.40318 | 0.906 |
| OGLE-BLG-CEP-02 | 6498.44740 | 0.914 |
| OGLE-BLG-CEP-03 | 6463.56319 | 0.545 |
| OGLE-BLG-CEP-05 | 6409.47279 | 0.845 |
| OGLE-BLG-CEP-32 | 6498.47404 | 0.727 |
| 2MASS J18181710-3401088 | 6498.48681 | (+35 kms$^{-1}$) |
| 2MASS J18182553-3349465 | 6498.50206 | (+0 kms$^{-1}$) |

For each object named, the heliocentric Julian date (HJD) when the spectrum was obtained and the phase of the Cepheid variations is listed. For the two reference stars, the catalogue radial velocities ($V_R$) are given.

Radial velocities were obtained by cross-correlation of the spectra with a synthetic spectrum taken from the library assembled for the RAVE experiment[34]. Two stars with known radial velocities[35] were used as a check on the radial velocity zero point. The measured velocities for these two stars are 34.7 km s$^{-1}$ and −12.0 km s$^{-1}$, respectively. Mean radial velocity errors due to photon statistics are 10 km s$^{-1}$, so the radial velocity zero point seems to be secure. The measured heliocentric velocities have been corrected for stellar pulsation adopting a standard velocity curve for short-period Cepheids with a semi-amplitude of 20 km s$^{-1}$ (for example, figure 6 of ref. 36). The mean heliocentric velocity after correction is 4 ± 8 km s$^{-1}$ compared with 13 ± 7 km s$^{-1}$ before correction. This indicates that uncertainties in the correction will not affect our conclusion regarding the mean radial outward velocity of this group of stars and that the error given in the main text is realistic.

**Galactic structure**
In the main paper and in the following we adopt a distance from the Sun to the Galactic centre of 8.5 kpc and a flat rotation curve with a velocity $\theta = 220$ km s$^{-1}$, to allow for a direct comparison with models describing the H I gas behaviour in the outer Galactic disk[2]. Plausible changes[37] in these values will not affect our conclusions.

The heliocentric distances of the Cepheids (*D* values in Table 1) are comparable with that of the Sgr dwarf galaxy (about 24 kpc), and a tidal stream from this system crosses the Galactic plane, behind the Galactic centre, close to the Galactic bulge at positive Galactic longitude. RR Lyrae variables belonging to this stream have recently been found in our field[38] at a distance of around 27 kpc. The possibility that the Sgr system contains stars as young as about 100 Myr has also been raised[39]. This is the expected age of short-period classical Cepheids[15], so we cannot rule out the possibility that our stars belong to the Sgr system on the basis of photometry alone, and kinematic information is essential. Because the heliocentric radial velocities of Sgr dwarf members are about 150 km s$^{-1}$ (refs 6 and 7), it is clear from the velocities in Table 1 that our Cepheids belong instead to the far outer parts of the Galactic disk. The possible association of the Cepheids with the far outer molecular spiral arm[5] was raised in the main paper. This arm lies at positive Galactic longitudes (in the first quadrant). At *l* = 13°.25, the lowest longitude at which the carbon monoxide (CO) was measured, the estimated distance[5] is $D = 23$ kpc, corresponding to $R = 14.5$ kpc; the exact distances are sensitive to the kinematic model. These are somewhat less than the $D$ and $R$ values in Table 1. Our values (in Table 1) of course refer to a region where there is no information from the gas. Adopting an alternative kinematic model[2] with elliptical orbits will lead to larger derived distances of the gas.

The five Cepheids are concentrated in a relatively small part of the area covered by the OGLE survey at positive longitudes. It is possible that variable interstellar absorption over the field could account for this. However, it seems more likely that it is due to real clumping, such as is common for young objects in spiral arms. The far outer molecular arm has not yet been seen emerging from the Galactic centre region at negative longitudes.

We note that our stars are at Galactocentric distances comparable to, but greater than, those of a small number of masers defining an outer arm in the general direction of the anti-centre[40]. The Cepheids whose radial velocities were studied in the general direction of the anti-centre are also at somewhat shorter distances (mean $R = 12.9$ kpc)[20].

The Sun–Cepheid–Galactic centre angle is small for all of our stars (0° to 2.7°, as measured in the Galactic plane). Thus, the corrected radial velocity, $\rho$, primarily measures motion that is radial from the Galactic centre and the five Cepheids give a mean $\rho = 23 \pm 9$ km s$^{-1}$, indicating a significant mean outward radial motion. This would not be inconsistent with a value of 9 km s$^{-1}$ predicted in the model[2]. It should also be noted that in the general solar neighbourhood systematic deviations from circular motion of about 10 km s$^{-1}$ are known for OB stars and Cepheids in regions of around a kiloparsec radius[41]. Therefore, our result does not necessarily imply any significant general deviation from circular orbits. The uncertainty in this conclusion is related to the small number of objects involved rather than to the uncertainties in estimating mean Cepheid velocities.

In the general anti-centre region at somewhat smaller Galactocentric distances no evidence was found[20] for a general outward velocity, though curiously the three Cepheids in that study[20] with Galactic latitudes of $|\Delta l| < 10°$ from the anti-centre have a mean positive radial velocity of $10 \pm 4$ km s$^{-1}$. Despite the small number of our stars the result would be in conflict with an outward motion of the local standard of rest of about 14 km s$^{-1}$ which has sometimes been suggested[42] to explain the Galactic asymmetry of the H I velocities.

**Number of Cepheids observed and expected**
With such a small number of Cepheids in our sample it is impossible to carry out a detailed study of the space distribution at their distance from the Galactic centre. However, the following rough calculation shows that their presence far from the Galactic plane requires the presence of a flared disk.

Consideration of the number of Cepheids in the solar neighbourhood suggests that the expected number of such stars in a column perpendicular to the Galactic plane and with a cross-sectional area of one square kiloparsec is about 60. With a scale height of 86 pc, as for the gas[2] (HWHM 60 pc) and taking into account that the area on the Galactic plane, between $D = 20$ kpc and $D = 30$ kpc, covered by the OGLE survey of (slightly less than) 44 kpc$^2$, the number of Cepheids expected in that survey with $z > 1$ kpc is approximately 0.008. This is for solar neighbourhood densities. With a disk scale length of 3 kpc, the drop in density from 8.5 kpc to 15 kpc is a factor of 9 and the expected number of Cepheids is about 0.001 (that is, for an unflared disk Cepheids are not expected). However, at the distance of our Cepheids the scale height of the gas is about 577 pc (HWHM 400 pc) and if the Cepheids follow the gas we predict the existence of about 18 in the relevant region. This calculation is obviously quite uncertain, but it is sufficient to show that our conclusion that these Cepheids are in the outer regions of a flared disk of scale height similar to that of the gas is entirely plausible. We see no other satisfactory explanation for these stars. We cannot rule out the possibility that a few more of the OGLE variables are classical Cepheids.

Owing to the small numbers, the likely effects of non-uniform interstellar absorption and the fact that young objects are expected to be found in groups rather than uniformly distributed over the field, it is not feasible to draw any strong conclusion from the fact that these Cepheids are confined to the positive longitude side of the OGLE field or that four of the five stars are at northern latitudes, despite the fewer OGLE fields there.

**Acknowledgments:** The observations reported in this Letter were obtained with the Infrared Survey Facility (IRSF) at Sutherland and the Southern African Large Telescope (SALT). MWF, JWM and PAW gratefully acknowledge research grants from the South African National Research Foundation.

**Affiliations**
1. Astrophysics, Cosmology and Gravity Centre, Astronomy Department, University of Cape Town, Rondebosch 7701, South Africa
   - Michael W. Feast & Patricia A. Whitelock
2. South African Astronomical Observatory, PO Box 9, Observatory 7935, South Africa
   - Michael W. Feast, John W. Menzies & Patricia A. Whitelock
3. Department of Astronomy, School of Science, The University of Tokyo, 7-3-1 Hongo, Bunkyo-ku, Tokyo 113-0033, Japan
   - Noriyuki Matsunaga

**Contributions**
M.W.F. coordinated the project and conducted the analysis. J.W.M. reduced the spectroscopic observations from SALT and determined the radial velocities. N.M. made and analysed the IRSF observations. All four authors contributed to the explanation and the discussion.